# 3D Gaussian Adaptive Reconstruction for Fourier Light-Field Microscopy

Chenyu Xu,[a,†] Zhouyu Jin,[b,c,†] Chengkang Shen,[a] Hao Zhu,[a] Zhan Ma,[a] Bo Xiong,[d,*] You Zhou,[b,c,*] Xun Cao,[a] Ning Gu[b,c]

[a] School of Electronic Science and Engineering, Nanjing University, Nanjing, Jiangsu, 210023, China
[b] Medical School, Nanjing University, Nanjing, Jiangsu, 210093, China
[c] Nanjing Key Laboratory for Cardiovascular Information and Health Engineering Medicine. Institute of Clinical Medicine, Nanjing Drum Tower Hospital, Medical School, Nanjing University, Nanjing, 210093, China
[d] National Key Laboratory for Multimedia Information Processing, Peking University, Beijing 100871, China
[†] These authors contributed equally to this work.

**Abstract**. Compared to light-field microscopy (LFM), which enables high-speed volumetric imaging but suffers from non-uniform spatial sampling, Fourier light-field microscopy (FLFM) introduces sub-aperture division at the pupil plane, thereby ensuring spatially invariant sampling and enhancing spatial resolution. Conventional FLFM reconstruction methods, such as Richardson-Lucy (RL) deconvolution, exhibit poor axial resolution and signal degradation due to the ill-posed nature of the inverse problem. While data-driven approaches enhance spatial resolution by leveraging high-quality paired datasets or imposing structural priors, Neural Radiance Fields (NeRF)-based methods employ physics-informed self-supervised learning to overcome these limitations, yet they are hindered by substantial computational costs and memory demands. Therefore, we propose 3D Gaussian Adaptive Tomography (3DGAT) for FLFM, a 3D gaussian splatting based self-supervised learning framework that significantly improves the volumetric reconstruction quality of FLFM while maintaining computational efficiency. Experimental results indicate that our approach achieves higher resolution and improved reconstruction accuracy, highlighting its potential to advance FLFM imaging and broaden its applications in 3D optical microscopy.

*Bo Xiong, E-mail: xiongbo@pku.edu.cn
*You Zhou, E-mail: zhouyou@nju.edu.cn

## 1 Introduction

Light-field microscopy (LFM)[1-4] has emerged in recent decades as an advanced optical imaging technique capable of capturing three-dimensional (3D) information in a single shot, enabling high-speed volumetric imaging while minimizing photobleaching and phototoxicity. However, its limited 3D spatial resolution and non-uniform spatial sampling scheme[5] have hindered its broader applicability. To overcome these limitations, Fourier light-field microscopy (FLFM)[6] has been developed, offering enhanced spatial resolution and spatially invariant sampling compared to conventional LFM. FLFM achieves multi-view imaging by dividing the sub-aperture at the pupil plane, thereby generating multiple parallax views on the camera sensor.

For 3D reconstruction in FLFM, the Richardson-Lucy (RL) deconvolution algorithm[7,8] and its variants are among the most widely used methods. However, due to the inherently ill-posed nature of the reconstruction problem, these approaches often suffer from poor axial resolution and significant signal loss, limiting FLFM's ability to generate high-quality volumetric data. Recently, data-driven reconstruction techniques for FLFM[9] have demonstrated notable improvements in spatial resolution. However, their performance relies on access to high-quality paired datasets or specific structural assumptions about the sample, which restricts their generalizability to diverse sample types or those with significant structural differences.

In contrast, Neural Radiance Fields (NeRF)-based methods, which employ self-supervised learning by integrating physical imaging models, have been successfully applied in optical microscopy imaging, such as Artefact-free Refractive-index Field (DeCAF)[10], Fourier Ptychographic Microscopy with Implicit Neural Representation (FPM-INR)[11], Computational Adaptive optics (CoCoA)[12], and volumetric wide-field microscopy with physics-informed ellipsoidal coordinate encoding implicit neural representation (PIECE-INR)[13]. While these methods achieve high-quality reconstructions without requiring large-scale paired training datasets, their substantial computational overhead and GPU memory costs caused by dense sampling strategy remain challenging[14].

Recently, 3D Gaussian Splatting (3DGS), an emerging multi-view 3D rendering technique, has shown exceptional performance in the field of computer vision, outperforming NeRF in computational efficiency while maintaining comparable rendering quality[15]. Furthermore, prior works have extended its use to volumetric reconstruction tasks, including CT reconstruction[16] and vessel reconstruction using digital subtraction angiography (DSA) images[17]. Its ability to model complex 3D structures with high accuracy makes it a promising candidate for extending its applications beyond computer vision to encompass the volumetric reconstruction of optical microscopy.

In this work, we propose 3D Gaussian Adaptive Tomography (3DGAT) as a 3D gaussian splatting based self-supervised learning method for 3D reconstruction of FLFM. By integrating efficient 3D Gaussian representation with the physical imaging model of FLFM, we demonstrate significant resolution improvement over conventional deconvolution-based methods. The main contribution of our 3DGAT relies on: (1) introducing the Gaussian-based representation into light microscopy and further applying it to FLFM modality for the first time; (2) implementing a robust initialization and loss constraint for the specific FLFM task; and (3) validating its effectiveness on both simulated data and real experimental dataset.

## 2 Methods

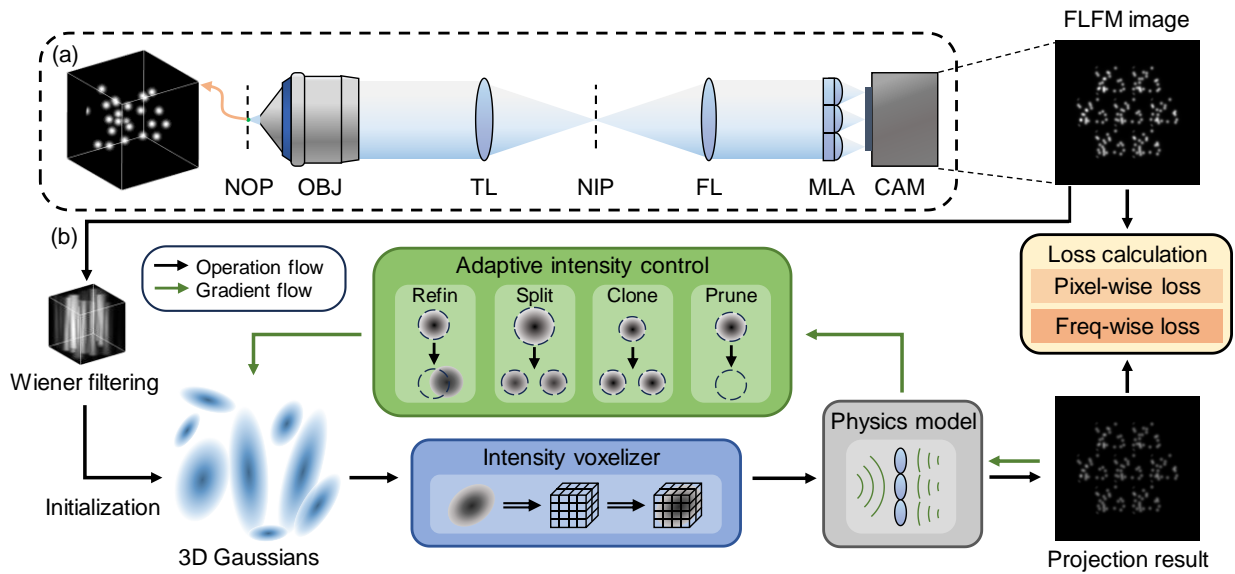

**Fig. 1. Principle of 3D Gaussian Adaptive Tomography (3DGAT).** (a) The optical set-up and forward imaging process of the FLFM system. (b) The schematic of the training pipeline of 3DGAT. NOP, native object plane; OBJ, objective lens; TL, tube lens; NIP, native image plane; FL, Fourier lens; MLA, microlens array; CAM, camera.

## 2.1 Forward imaging model of FLFM

By positioning a microlens array (MLA) at the pupil plane of the objective lens in an inverted fluorescence microscope, FLFM enables multi-view imaging of the observed object, generating multiple parallax views on the camera sensor, as illustrated in Fig. 1(a). Accordingly, FLFM can be approximated as a linear system, where the 3D spatial distribution of the fluorescence signal within the object space $O(r_o)$ is sampled, propagates through the optical system, and is ultimately projected onto the camera sensor.

Therefore, the physical imaging model of FLFM can be described as the convolution of object's distribution $O(r_o)$ and the point spread function (PSF) of the FLFM system $h(r_o, r_i)$:

$$I(r_i) = \sum_z \int O(r_o; z) h(r_o; z, r_i) dr_o, \tag{1}$$

where $r_o = (x_o, y_o) \in \mathbb{R}^2, z \in \mathbb{R}$ represents the object space coordinates, and $r_i = (x_i, y_i) \in \mathbb{R}^2$ is the sensor plane coordinates.

The detailed computation of $h(r_o; z, r_i)$ can be referred to a previous work by W. Liu et al.[18]. For numerical computations, $O(r_o; z)$ and $h(r_o; z, r_i)$ are discretized, enabling the relationship between the object volume $O$ and the captured image $I$ to be formulated as a discrete convolution:

$$I = \sum_j H_j * O_j, \tag{2}$$

where $H_j$ is the discretized format of $h(r_o; r_i)|_{z=j}$ and $*$ denotes the discrete 2D convolution operation. Therefore, the process of FLFM reconstruction involves solving an inverse problem for $O$, using the measured image $I$ along with the PSF $H$, which is either simulated based on system parameters or experimentally obtained as the measured PSF.

## 2.2 Realization of 3DGAT method

To solve this problem, we represent the target object $\widehat{\boldsymbol{O}}$ with a set of 3D Gaussian kernels $\mathbb{G}^3 = \{G_i\}_{i=1}^m$, each kernel defines a Gaussian-shaped fluorescence intensity field in 3D space:

$$G_i(\mathbf{x}|\rho_i, \boldsymbol{\mu}_i, \boldsymbol{\Sigma}_i) = \rho_i \cdot \exp\left(-\frac{1}{2}(\mathbf{x}-\boldsymbol{\mu}_i)^{\mathrm{T}}\boldsymbol{\Sigma}_i^{-1}(\mathbf{x}-\boldsymbol{\mu}_i)\right), \tag{3}$$

where $\rho_i$, $\boldsymbol{\mu}_i$ and $\boldsymbol{\Sigma}_i$ are learnable parameters, representing the central density, mean, and covariance of a 3D Gaussian ellipsoid. Considering the physical meaning of these parameters, $\boldsymbol{\Sigma}_i$ can be decomposed into scaling matrix $\boldsymbol{S}_i$ and rotation matrix $\boldsymbol{R}_i$, which are further represented as scaling vector $\boldsymbol{s}_i$ and rotation quaternion $\boldsymbol{r}_i$. These parameters control the ellipsoid's fluorescence intensity, central position, and scaling/rotation relative to a standard sphere in 3D space. In other words, $\rho_i$, $\boldsymbol{\mu}_i$, and $\boldsymbol{\Sigma}_i = \boldsymbol{R}_i\boldsymbol{S}_i\boldsymbol{S}_i^T\boldsymbol{R}_i^T$ can be used to generate arbitrary ellipsoids in 3D space.

In our proposed 3DGAT method, we first generate a coarse reconstruction by applying Wiener filtering to the raw FLFM measurement[19], which is formulated as:

$$\boldsymbol{O}_{\text{wiener\_filter}}(\boldsymbol{r}_o; z) = \mathcal{F}^{-1}\left\{\frac{\tilde{I}(\boldsymbol{r}_i)\widetilde{H}^*(\boldsymbol{r}_i; z, \boldsymbol{r}_o)}{\left|\widetilde{H}(\boldsymbol{r}_i; z, \boldsymbol{r}_o)\right|^2 + w^2}\right\}, \tag{4}$$

where $\tilde{\cdot}$ denotes the 2D Fourier transform, $\mathcal{F}^{-1}\{\cdot\}$ denotes the 2D inverse Fourier transform, $\boldsymbol{O}_{\text{wiener\_filter}}$ denotes the filtered result, and $w$ is the Wiener parameter. Subsequently, we sample from this preliminary reconstruction result to generate a set of 3D Gaussian kernels, as a robust initialization approach of our Gaussian-based reconstruction framework.

Inspired by the tile-based rasterizer on novel view synthesis tasks[15], we customize our intensity voxelizer based on the R$^2$-Gaussians method[16] to enable an efficient transformation from Gaussian-based to voxel-based representation. The voxelized data are then projected through the FLFM system's physics model to produce the final imaging results, as depicted in Eq. (2).

After this step, we introduce and adopt an objective loss function to mitigate blurring and artifacts during reconstruction, which follows the below formulation:

$$\mathcal{L} = \mathcal{L}_{\text{MSE}}(I_{\text{proj}}, I) + \alpha \mathcal{L}_{\text{FDL}}(I_{\text{proj}}, I), \tag{5}$$

where $I_{\text{proj}}$ denotes the projected results, and $I$ denotes raw FLFM measurement. The loss function in use constructs with two terms, the Mean Square Error (MSE) in space domain $\mathcal{L}_{\text{MSE}}$ to keep basic data similarity, and the Fourier Domain Loss (FDL) $\mathcal{L}_{\text{FDL}}$ to ensure further details alignment in frequency domain, which is defined as $\mathcal{L}_{\text{FDL}}(x, y) = \sum |\tilde{x} - \tilde{y}|$. $\alpha$ is a weight parameter, which empirically takes around $10^{-3}$.

On the optimization step, the gradient of each gaussian kernels is automatically back-propagated and accumulated owing to the differentiability of the intensity voxelizer and the physics model. Then four different strategies named "refine", "split", "clone" and "prune" are adopted to optimize the 3D Gaussian-based fluorescence intensity field[15,16].

*2.3 Computation details*

3DGAT is trained on a machine equipped with AMD EPYC 9654 processor and RTX 4090 graphic cards (NVIDIA). The software environment consists of Python 3.9.21 and Pytorch 2.1.1 with CUDA 12.1 support. To leverage GPU acceleration, the tile-based intensity voxelizer is developed by customizing CUDA kernel, building upon previous frameworks[15,16].

# 3 Results

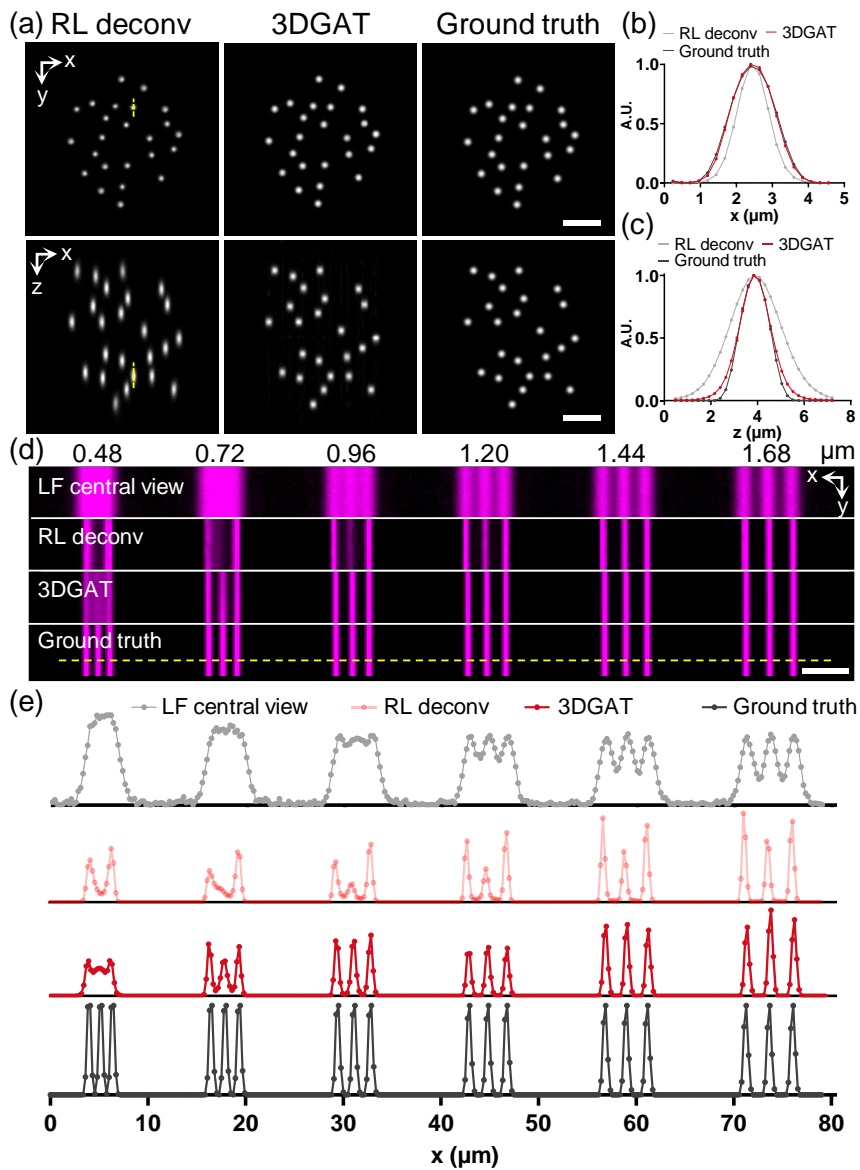

**Fig. 2. Performance and resolution evaluation of 3DGAT on synthetic data.** (a) Maximum intensity projections (MIPs) of synthetic fluorescent beads restored by RL deconvolution and 3DGAT, compared to the ground truth. (b-c) Intensity profile comparisons of RL deconvolution (gray) and 3DGAT (red) with the ground truth (black) along the yellow dashed lines in (a). (d) MIPs of synthetic fluorescent lines with varying intervals at the LF central view, along with RL deconvolution, 3DGAT, and the ground truth. (e) Intensity profile comparisons of the LF central view (gray), RL deconvolution (pink), and 3DGAT (red) with the ground truth (black) along the yellow dashed lines in (d). Scale bar: (a) 15μm; (d) 5μm.

To evaluate the performance and resolution of 3DGAT, we first use synthetically generated isotropic 3D fluorescent beads as the ground truth. Using the wave optics model of FLFM[18], we project the 3D images into 2D Fourier light-field images, with the FLFM parameters set to seven perspective views and a 20×/0.45 NA objective lens. The 3D reconstruction performance of 3DGAT is compared with RL deconvolution applying 100 iterations. As shown in Fig. 2(a), the results solved by 3DGAT are in good agreement with the ground truth. However, the RL deconvolution resolves the beads too finely in the lateral direction, while elongating it in the axial direction, as the line profiles shown in Fig. 2(b-c).

To further test the lateral resolution of 3DGAT, we follow the Rayleigh criterion and generate fluorescent lines with interline spacing from 1.68 µm to 0.48 µm and a line width of 0.24 µm. The central views of the projected light-field images and the restored results obtained using different methods are presented in Fig. 2(d). We also analyze and show the intensity profiles along the lines indicated by the yellow dashes in Fig. 2(d). As shown in Fig. 2(e), an obscure line in the light-field image could be clearly resolved as three parallel lines 0.72 µm apart using our proposed 3DGAT method. When the line-pair spacing is 0.48 µm, the lines could still be faintly discerned by 3DGAT. Meanwhile, the RL deconvolution method can barely resolve lines spaced 0.96 µm apart.

The simulation results clearly demonstrate that, compared to traditional RL deconvolution, the proposed 3DGAT method achieves higher-fidelity reconstruction with improved resolution. It effectively mitigates overfitting in the lateral direction while enhancing resolution in the axial direction.

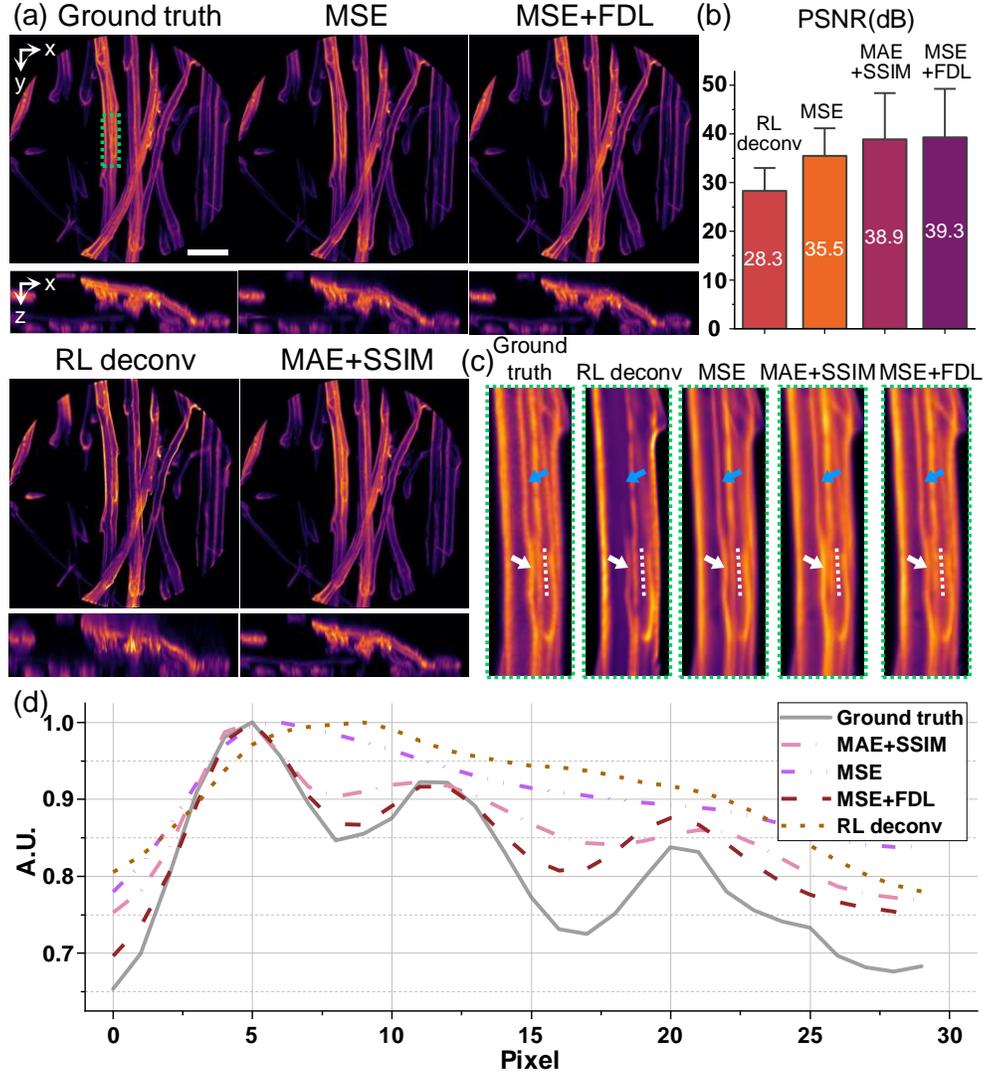

**Fig. 3. Comparison between RL deconvolution and 3DGAT with different loss functions.** (a) MIPs of simulated dandelion sample restored by RL deconvolution and 3DGAT with different loss (MSE, MAE+SSIM, and MSE+FDL loss), comparing with the ground truth. (b) Peak signal-to-noise ratios (PSNRs) of different methods across the depth range. (c) Enlarged views of the green dashed box in (a). (d) Normalized intensity profiles of different methods along the white dashed line in (c). Scale bar: 100μm.

To validate the effectiveness of our method on complex samples, we conduct a comparative simulation using a dandelion sample, utilizing FLFM with seven perspective views and a 20×/0.45 NA objective lens. We compare the results of RL deconvolution (20 iterations for the best performance and highest PSNR value) and the results obtained using 3DGAT with different loss functions (MSE, MAE+SSIM, and our proposed MSE+FDL).

The MIP images from x-y and x-z two direction of the reconstruction results, as well as the ground truth, are presented in Fig. 3(a). The results of 3DGAT have shown better optical sectioning ability in Fig. 3(a) with more details in x-y plane than RL deconvolution. This can better be observed in the enlarged images in Fig. 3(c), pointed by the blue arrows. The quantitative layer-wise PSNRs in Fig. 3(b) further confirm this observation.

Furthermore, we evaluate the performance of 3DGAT with different loss functions, such as the MAE+SSIM loss from the original 3DGS paper[15], the commonly used MSE loss for reconstruction tasks, and our proposed MSE+FDL loss. In the enlarged view in Fig. 3(c), our proposed loss demonstrated better visual quality than others when comparing with the ground truth (details pointed by the white arrow). The normalized intensity profile along the white dashed line in Fig. 3(d) and the PSNR metrics in Fig. 3(b) also support this finding. A close examination of the profile in Fig. 3(d) reveals that our proposed MSE+FDL loss achieves optimal alignment with the ground truth, accurately reproducing three distinct peaks at identical positions.

This simulation demonstrates that the proposed 3DGAT method outperforms traditional RL deconvolution when applied to complex biological samples, offering both enhanced resolution and improved retention of fine details. We also validate that the proposed spatial-frequency domain simultaneous constraint loss (MSE+FDL loss) more accurately reconstructs the 3D structural information of the sample compared to existing loss functions.

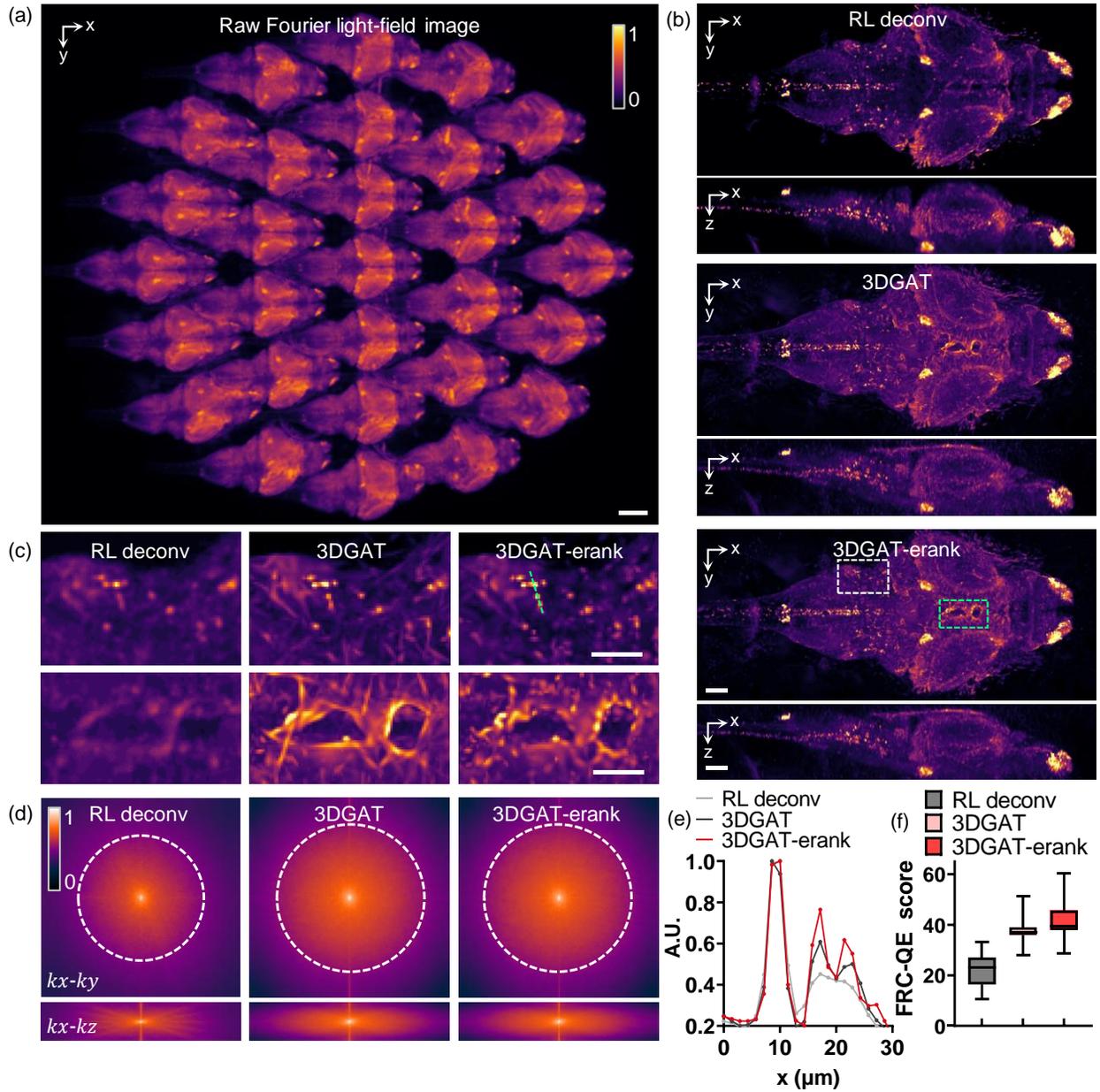

**Fig. 4. Reconstruction on experimentally captured zebrafish data.** (a) Raw Fourier light-field image of zebrafish data acquired from [20]. (b) x-y MIP images of results obtained by RL deconvolution, raw 3DGAT and effective-rank-regularized 3DGAT (3DGAT-erank). (c) Enlarged view of the white and green dashed boxes in (b). (d) Fourier domain visualization of x-y and x-z MIP images recovered by the corresponding methods. (e) Normalized intensity profile along the green dashed line in (c). (f) Fourier ring correlation quality estimate (FRC-QE) scores of three methods, respectively. Scale bar: 50μm (a, b), 30μm (c).

Finally, we validate our method using real experimental zebrafish data. The data[20] is acquired using an FLFM system with 29 views and a 16×/0.8NA water-immersion objective lens, as shown in Fig. 4(a). We further incorporate the effective rank regularization[21] to eliminate the needle-like artifacts caused by noise in the experimentally captured images. MIP images obtained using 3DGAT with (3DGAT-erank) and without (3DGAT) effective rank regularization, along with RL deconvolution, are displayed together in Fig. 4(b). The enlarged views in Fig. 4(c) demonstrate that the 3DGAT-erank preserves more structural details than RL deconvolution while exhibiting fewer needle-like artifacts than the raw 3DGAT. This observation is further supported by the intensity profiles shown in Fig. 4(e). The Fourier domain visualization of the x-y and x-z MIP images reconstructed by the respective methods in Fig. 4(d) highlights the effectiveness and high resolution of our proposed method. Additionally, we compute and present the Fourier ring correlation quality estimate (FRC-QE) scores in Fig. 4(f), where a higher score indicates superior recovery of frequency details. This experiment validates the effectiveness of our method using real zebrafish data, demonstrating that incorporating effective rank regularization enhances structural detail preservation while reducing needle-like artifacts.

## 4  Conclusion

In conclusion, we introduce 3DGS into optical microscopy imaging for the first time and apply it to FLFM, proposing a novel 3D Gaussian Adaptive Tomography (3DGAT) framework. By leveraging robust initialization and loss constraints, while integrating the physical imaging model with efficient 3D Gaussian representations, our method enables high-quality 3D fluorescence reconstruction with adaptive intensity adjustment. In the future, we plan to integrate the physical models of various multi-view acquisition schemes in optical microscopy into our proposed 3DGS-based framework, such as multi-view light-sheet microscopy[22,23] and others advanced techniques[24,25], while incorporating efficient regularization derived from corresponding physical priors and constraints, thereby balancing 3D reconstruction performance, data requirements, and computational cost. Additionally, in our current implementation of 3DGAT, we voxelize the 3D Gaussian distribution after its computation and then numerically simulate the physical processes to estimate the measurements. This step undeniably introduces additional computational overhead and time compared to the original analytic approach for the 3DGS. Therefore, we seek to develop analytic mathematical representations of the imaging process across different microscopy modalities[26], mapping the 3D distribution to 2D acquisition, to further enhance the efficiency and accuracy of the 3DGS-based method in this field. We believe this work establishes a valuable foundation for extending 3DGS to a broader range of microscopic imaging applications.

*Disclosures*

The authors declare no conflicts of interest.

*Acknowledgments*

This work was supported by the National Key Research and Development Program of China (2024YFF0508604); the Natural Science Foundation of Jiangsu Province (BK20222002); and the National Natural Science Foundation of China (62071219, 62025108, 62371006).

**Caption List**

**Fig. 1** Principle of 3D Gaussian Adaptive Tomography (3DGAT).

**Fig. 2** Performance and resolution evaluation of 3DGAT on synthetic data.

**Fig. 3** Comparison between RL deconvolution and 3DGAT with different loss functions.

**Fig. 4** Reconstruction on experimentally captured zebrafish data.